\documentclass[journal=jacsat,manuscript=article]{achemso}

\usepackage[version=3]{mhchem} 
\usepackage{graphicx}
\usepackage{amssymb}
\usepackage{lineno}
\usepackage{comment}
\usepackage{amsmath}
\usepackage{tabu}
\usepackage{xcolor}
\usepackage{setspace}
\usepackage{bm}
\usepackage{xcolor}
\usepackage{multirow} 
\usepackage{array}
\usepackage{booktabs} 
\usepackage[group-separator={,}]{siunitx} 
\usepackage{subcaption}

\newcommand{\add}[1]{\textcolor{black}{#1}}

\author{Xianglin Liu}
\email{xianglinliu01@gmail.com}
\affiliation[Pengcheng Laboratory]
{Pengcheng Laboratory, Shenzhen, 518000, China}
\altaffiliation{These authors contributed equally.}

\author{Kai Yang}
\affiliation[Pengcheng Laboratory]
{Pengcheng Laboratory, Shenzhen, 518000, China}
\altaffiliation{These authors contributed equally.}

\author{Fanli Zhou}
\affiliation[Xiangnan University]
{School of Computer and Artificial Intelligence, Xiangnan University, Chenzhou, 423000, China}
\altaffiliation{These authors contributed equally.}

\author{Pengxiang Xu}
\affiliation[Pengcheng Laboratory]
{Pengcheng Laboratory, Shenzhen, 518000, China}

\title[SMC=X]{SMC-X: A Distributed Scalable Monte Carlo Simulation Method for Chemically Complex Alloys}

\abbreviations{IR,NMR,UV}
\keywords{American Chemical Society, \LaTeX}

\begin{document}

\begin{tocentry}

\includegraphics[width=0.6\linewidth]{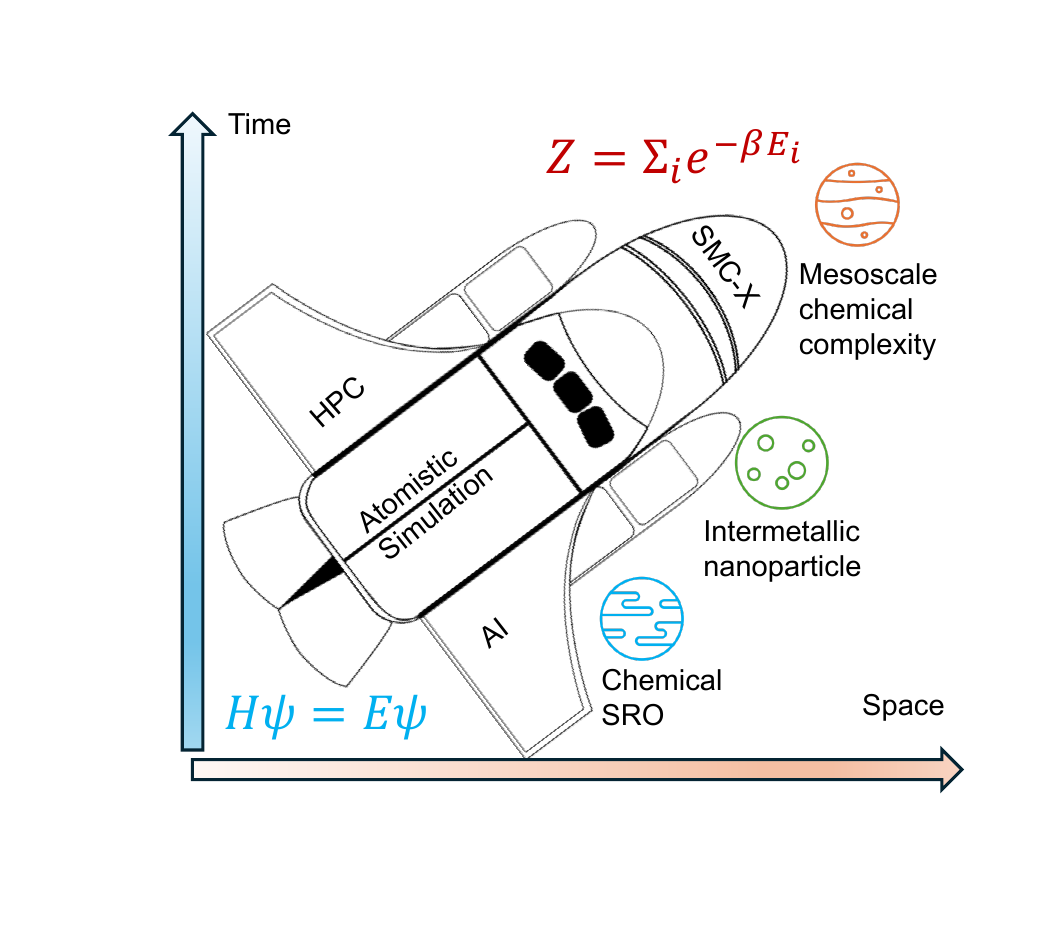}

\end{tocentry}

\begin{abstract}
    To predict the complex chemical evolution in multicomponent alloys, it is highly desirable to have accurate atomistic simulation methods capable of reaching sufficiently large spatial and temporal scales. In this work, we advance the recently proposed SMC-X method through distributed computation on either GPUs or CPUs, pushing both spatial and temporal scales of atomistic simulation of chemically complex alloys to previously inaccessible scales. This includes a \add{record-breaking 128-billion-atom} HEA system extending to the micrometer regime in space, and a 1-billion-atom HEA evolved over more than three million Monte Carlo swap steps, approaching the minute regime in time. We show that such large-scale simulations are essential for bridging the gap between experimental observations and theoretical predictions of the nanoprecipitate sizes in HEAs, based on analysis using the Lifshitz–Slyozov–Wagner (LSW) theory for diffusion-controlled coarsening. This work demonstrates the great potential of SMC-X for simulation-driven exploration of the chemical complexity in high-entropy materials at large spatial and temporal scales.
\end{abstract}

\section{Introduction}
Materials are composed of atoms, and the types of atoms together with their spatial arrangements govern a wide spectrum of physical properties, including mechanical, electrical, magnetic, and thermal behavior. However, the vast chemical complexity present in many technologically important materials poses formidable challenges for existing computational approaches. This challenge is particularly pronounced in high-entropy alloys (HEAs) \cite{EGeorge_Nature}, which contain multiple principal elements and therefore exhibit substantially increased chemical complexity. Moreover, this complexity is further compounded by the interplay and competition between order and disorder, as well as enthalpic and entropic effects, across diverse spatial, temporal, and temperature scales, making predictive modeling of such systems exceptionally demanding.
Capturing these complexities, including chemical composition \cite{ChemicalComposition2019Nature}, disorder \cite{AtomComplexity}, chemical short-range order (SRO) \cite{SRO_NCS_2023, APT_CoCrNi_NatureMat_2024}, chemical medium-range order (MRO) \cite{mediumRangeOrderNC}, multi-component intermetallic nanoparticles (MCINP) \cite{Yang933}, and nano lamellae\cite{CTLiuNC, MaxPlankNC2025}, requires modeling and simulating atomic systems over large spatial and temporal scales, with accuracy and generalizability of first principles method \cite{LIU2023101018, LIU2019107955, Eisenbach_2019}. 

In recent years, machine learning potentials (MLPs) \cite{JCTCBP, Zhang_2025, MPNNs_2023} trained on DFT datasets \cite{OQMD, Horton2025} have emerged as a highly promising solution to the aforementioned bottlenecks \cite{UniversalModel_NCS, 2025Nature}.  By fitting interatomic potential energy surfaces with machine learning models, MLPs can retain near-DFT accuracy \cite{Ong_JPC_2020} while dramatically increasing simulation efficiency and extending accessible system sizes \cite{doi:10.1126/science.abn3445}. With an appropriate interaction cut-off radius, many MLPs can reduce the computational complexity of traditional DFT calculations from cubic scaling, $O(N^3)$, to linear scaling, $O(N)$ \cite{MPNNs_2023}, with respect to the number of atoms $N$. Moreover, by bypassing the explicit solution of the electronic structure, the computational prefactor can be reduced by orders of magnitude \cite{liuNPJ2025}, enabling simulations at spatial scales previously deemed intractable \cite{10.1145/3458817.3487400}. The integration of MLPs with molecular dynamics (MD), a foundational atomistic simulation technique renowned for its excellent scaling, has revolutionized computational materials science \cite{2025Nature}, chemistry \cite{Zhang2024}, and biology \cite{Gkeka2020} by enabling accurate, large-scale simulations at significantly reduced cost.

Although the MLP + MD approach successfully scales simulations to unprecedented sizes \cite{10.1145/3581784.3627041, 2020NPJ_Li, Yin2021}, its intrinsic limitations restrict accessible time scales to the nanosecond regime, which typically requires millions of sequential MD steps. Consequently, it is extremely challenging, if not impossible, for MD to capture many important slow-evolving phenomena, such as precipitate formation in alloys \cite{Yang933}. For instance, a single swap step between two neighboring lattice sites may take millions or even billions of MD steps to simulate. For such a purpose, Monte Carlo (MC) simulations offer unique advantages due to their characteristic of directly sampling configurational space, rather than tracking real-time trajectories as in MD \cite{Korman_npj, LIU2021110135}. 
However, compared to MD, the integration of conventional MC with MLP faces difficulties when scaled up to large systems \cite{liuNPJ2025}. This challenge stems from the intrinsically sequential update nature in mainstream Monte Carlo algorithms, such as the Metropolis method, where trial moves are proposed site by site, inherently limiting parallelization \cite{Anderson2016SMC}. This bottleneck is in stark contrast to that in MD, where all atoms can be updated simultaneously at each step.


Previous efforts to scale Monte Carlo (MC) simulations have introduced parallel trial moves using generalized checkerboard algorithms, with most implementations customized for specific interaction types or hardware \cite{PREIS20094468, BLOCK20101549, ROMERO2020107473, liu2019high, OrtegaZamorano2013FPGAIsing, Anderson2016SMC}, and only a few designed for general interactions \cite{sadigh2012scalable, liuNPJ2025}. Among all these implementations, the recently proposed scalable Monte Carlo at eXtreme (SMC-X) \cite{liuNPJ2025} method stands out due to its superb efficiency. Such a high efficiency is achieved via exposing the previously hidden degree of parallelism by dynamical link-cell (LC) decomposition and local interaction zone (LIZ),  and by carefully mapping these degrees of parallelisms onto modern GPU architectures to remove unnecessary data movement.
These innovations allow SMC-X to simulate systems as large as one billion atoms using a single GPU \cite{liuNPJ2025}. This performance not only substantially exceeds all prior MC + MLP methods, but also exhibits advantages against state-of-the-art (SOTA) MD + MLP implementations in both the number of atoms and the atomistic simulation throughput, measured by
\(N_{\rm atom} \times N_{\rm steps}\) per second
\cite{liuNPJ2025}. A comparison of various computational methods for chemically complex alloys is shown in Tab.~\ref{tab:methods}. It should be noted that while mesoscale methods such as the Potts model, kinetic Monte Carlo, and phase-fild methods can simulate very large spatial and temporal scales, they introduce coarse-grained approximations that rely on empirical parameters, making them less rigorous at the atomistic level, therefore are \add{not listed} in Tab.~\ref{tab:methods}.

\begin{table}[!ht]
\centering
\caption{Comparison of computational methods for studying chemically complex alloys at atomic scale. The temporal scale for MC method is estimated by multiple the estimated time for one swap step (1 $\mu$s) with the number of MC steps ($10^4$ for SPMC, and $10^6$ for SMC-X). Note that time estimation can vary significantly since it depends exponentially on temperature and the energy barrier.} \label{tab:methods}
\begin{tabular}{p{0.12\textwidth} p{0.1\textwidth} p{0.06\textwidth} p{0.3\textwidth} p{0.3\textwidth}}
\toprule
Method & Length  & Time & Primary Strength/ Limit & Typical Applications \\
\midrule
DFT & \AA--nm  & ps & High accuracy and generalizability, but computationally expensive &Phonon dispersion, stacking fault energies \\
EAM+MD & {nm--$\mu$m} & ns  & Highly efficient, but limited in accuracy and generalizability.& Plastic deformation simulation, radiation damage \\
MLP+MD &  nm--$\mu$m & ns & Excellent combination of accuracy, generalizability, and efficiency, but limited time-scale. & Plastic deformation simulation, general-purpose foundation models \\
MLP + Hybrid MC/MD & nm & ms & Can handle both structural relaxation and chemical complexity, but lost true kinetics.  & chemical SRO, GB segregation \\
MLP + SMC-X & nm–$\mu$m & s & Handle chemical complexity at large spatial and temporal scales, but no kinetics & Chemical nano- or micro- structures, e.g. precipitate\\
\bottomrule
\end{tabular}
\end{table}

\begin{figure} [ht!]
    \centering
    \includegraphics[width=0.8 \linewidth]{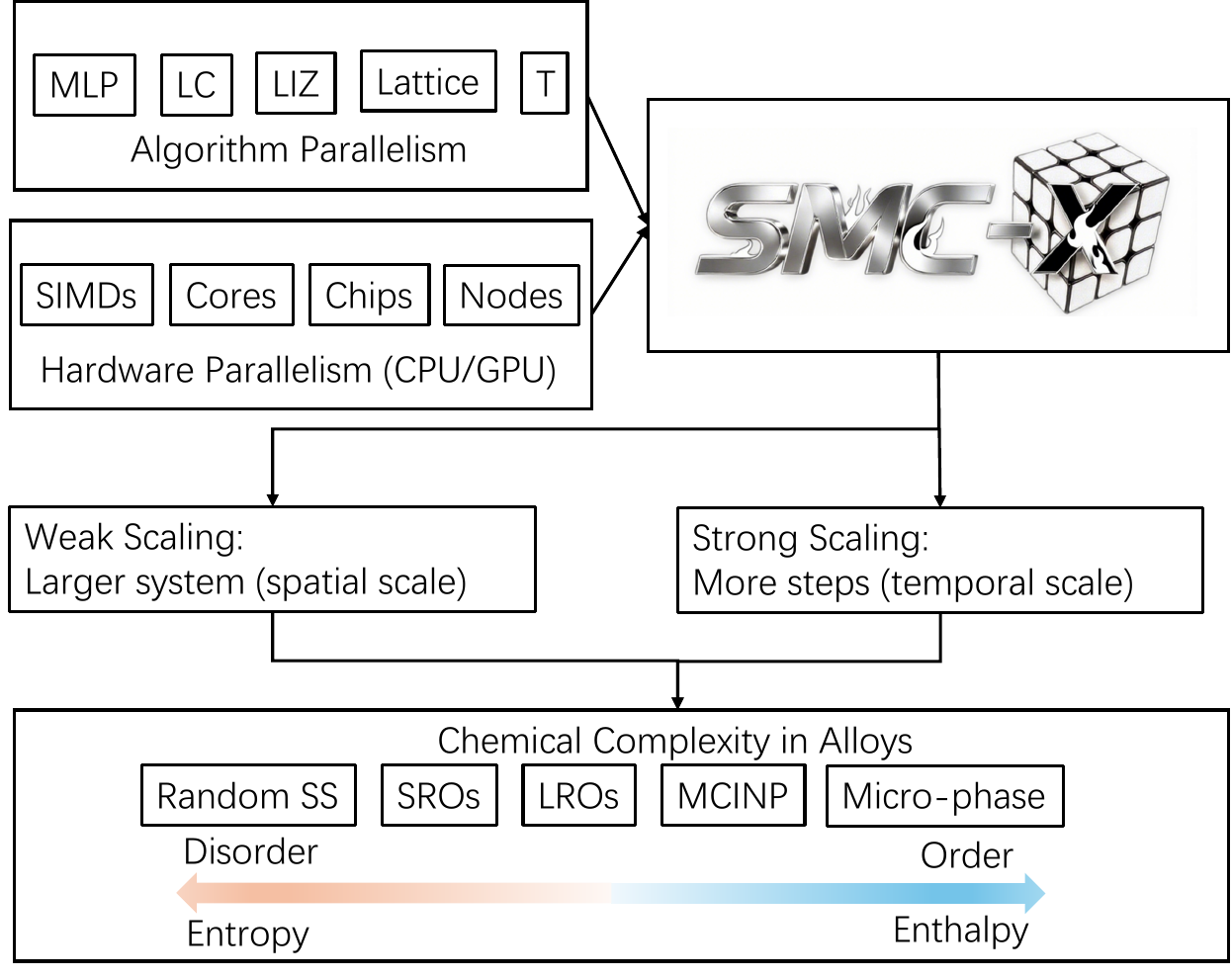}
    \caption{A schematic of the workflow to illustrate the main contributions. MLP: machine learning potential; LC: link-cell; LIZ: local interaction zone; T: temperature; SIMD: single-instruction multiple-data; SS: solid solution; SRO: short-range order; LRO: long-range order; MCINP: multicomponent intermetallic nanoparticle.}
    \label{fig:parallelism}
\end{figure}

In this work, we advance the SMC-X method to even larger spatial and temporal scales through distributed computing. We demonstrate that, by carefully mapping the algorithmic parallelism in SMC-X into the hardware parallelism in modern computational systems, we can achieve atom-by-atom simulation of chemically complex alloys at unprecedented spatial and temporal scales. We further demonstrate the versatility of SMC-X across different computation platforms with the efficient implementations on both the CPUs and GPUs. Collectively, these advances establish SMC-X as a framework that pushes the spatial and temporal scales of atomistic simulation to unparalleled levels, providing a crucial bridge between experiment and theory, as exemplified by the prediction of nanoprecipitate size in high-entropy alloy. An illustration of the method in this work is shown in Fig.~\ref{fig:parallelism}.

\section{Theory and Methods}
In this section, we present an introduction of the theoretical background of distributed SMC-X. Note that the basic concept of the SMC-X algorithm has been introduced in Ref.~\cite{liuNPJ2025}, but the focus was on its application in materials science. Here we present a more detailed discussion and introduce the distributed implementation on both CPUs and GPUs.
\subsection{Monte Carlo simulation}
In MC, physical observables are computed as ensemble averages over sampled configurations, providing direct access to thermodynamic properties. For instance, in the canonical ensemble (NVT), the probability of observing a configuration $\boldsymbol{\sigma}_k$ is given by:
\begin{align}
    P(\boldsymbol{\sigma}_k) = \frac{e^{-\beta H(\boldsymbol{\sigma}_k)}}{\sum_{\boldsymbol{\sigma}_k} e^{-\beta H(\boldsymbol{\sigma}_k)}}, \label{probability}
\end{align}
where $\beta = 1/(k_B T)$,  $k_B$ is the  Boltzmann constant, and $T$ is the temperature. Physical observables $\langle \hat{O} \rangle$ such as the expected energy, magnetization, specific heat, and magnetic susceptibility can then be calculated accordingly from
\begin{align}
    \langle \hat{O} \rangle = \sum_k \frac{ O(\sigma_k)e^{-\beta H(\boldsymbol{\sigma}_k)}}{e^{-\beta H(\boldsymbol{\sigma}_k)}}.
\end{align}
A sample following the probability distribution of Eq.~\ref{probability} can be efficiently generated via a Markov chain Monte Carlo (MCMC) simulation. For such a purpose, the Metropolis-Hastings algorithms is the most commonly used method. The Metropolis-Hasting algorithm starts with proposing a move (trial) at each lattice site, with the acceptance probability of this trial given by:
\begin{align}
P=\begin{cases}
1, &\Delta E \leq 0\\
\exp(-\beta{\Delta E}), &\Delta E >0.
\end{cases}
\end{align}
A Monte Carlo step, or sweep, is defined as attempting exactly one trial move per lattice site. An MC simulation typically begins with a series of warm‑up steps to ensure thermal equilibrium. After equilibration, the configurations are measured in subsequent steps, allowing the recorded samples to be used to estimate physical observables via statistics.

The trial moves within the Metropolis-Hasting algorithm are inherently sequential, which poses a significant challenge to parallelization. Although advanced techniques such as cluster MC \cite{ PhysRevLett.58.86} or the Wang–Landau algorithms \cite{PhysRevLett.86.2050, 10.1145/1654059.1654125, PhysRevE.97.043301} can alleviate this limitation by reducing or avoiding sequential site-by-site updating, they are often restricted in scope and lack general applicability. A more general approach leverages the assumption of short-range interactions: If interactions are sufficiently localized, then spatially distant regions can be updated independently and in parallel. In the simple Ising model, this principle underlies the well-known checkerboard algorithm. Extensions of this approach to three-dimensional lattices and beyond nearest-neighbor interactions are conceptually straightforward for specific models \cite{PREIS20094468}. However, generalization to arbitrary short-range interactions is more limited. One notable example is the Scalable Parallel Monte Carlo (SPMC) algorithm \cite{sadigh2012scalable}, as implemented in the LAMMPS package \cite{THOMPSON2022108171}. Despite its utility, SPMC has disadvantage for further scaling-up: to minimize the weak spatial correlations in SPMC algorithm, the total volume of the domains need to be maximized \cite{sadigh2012scalable}, which underutilizes the available parallelism, especially on architectures with a high degree of concurrency. As a result, current applications of SPMC are limited systems of less than a million atoms \cite{Yin2021, YAO2024120457}.

\begin{figure} [ht!]
    \centering
    \includegraphics[width=\linewidth]{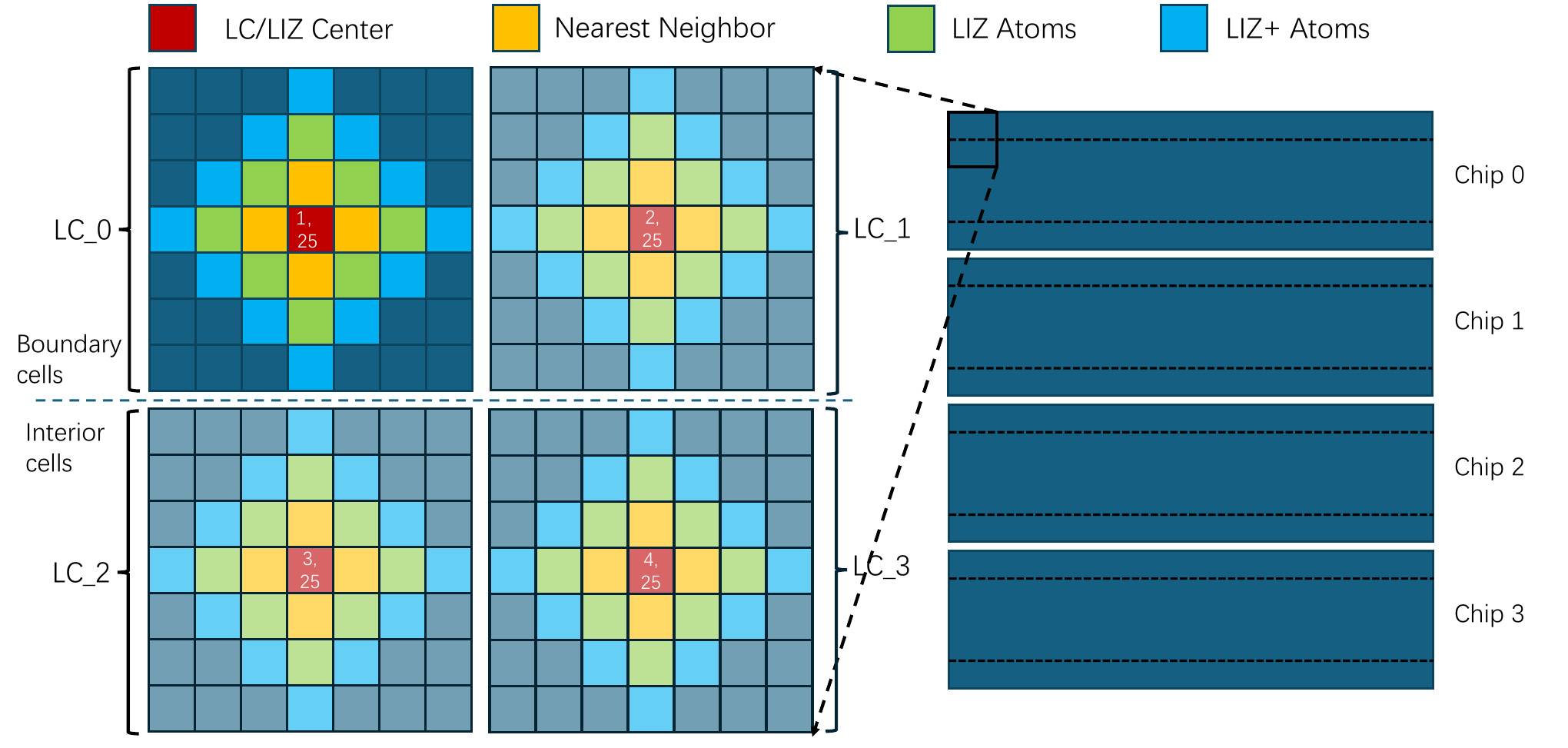}
    \caption{A schematic to illustrate the SMC-X algorithm, including link-cell (LC), local interaction zone (LIZ), and domain decomposition scheme, using the 2D square lattice as an example. LIZ+ represents sites needed for calculating the energy change due to possible MC trial. Adapted from Ref. ~\cite{liuNPJ2025}}
    \label{fig:SMC-X}
\end{figure}

\subsection{Distributed SMC-X} \label{DistributedSMC-X}
In the SMC-X framework \cite{liuNPJ2025}, the challenge of site-by-site sequential updating is addressed through two key innovations: dynamical link-cell (LC) and local interaction zone (LIZ), as illustrated in Fig.~\ref{fig:SMC-X}. Compared to SPMC \cite{sadigh2012scalable}, the LCs are dynamically constructed to ensure that every site is visited exactly once per MC step, and that the detailed balance condition is strictly satisfied. The resulting degree of parallelism is given by:
    \begin{equation}
        P_{LC} = N_{atom}/N_{LC},
    \end{equation}
where $N_{atom}$ is the total number of atoms in the supercell, and $N_{LC}$ is the number of atoms within LC. By choosing the smallest LC size that fully separates short-range interactions, the parallelism $P_{\mathrm{LC}}$ can be maximized. The LIZ is introduced to exploit the short-range interaction to reduce the computational complexity of an MC trial move from $O(N)$ to $O(1)$. Furthermore, by computing the atomic features within each LIZ on the fly and tightly coupling it with the MC simulation, the memory usage in SMC-X can be drastically reduced, and near-core caches can be effectively utilized to significantly enhance computational throughput.

In addition to LC and LIZ, there are additional parallel opportunities in SMC-X, as illustrated in the algorithm part in Fig.~\ref{fig:parallelism}. Among them, the lattice and the temperature parallelism can be implemented as MPI ranks for distributed computation over multiple chips. In lattice parallelism, the complete lattice is partitioned across multiple GPUs, with a ghost layer used to exchange boundary information between neighboring sublattices, as illustrated in Fig.~\ref{fig:SMC-X}. Under this scheme, if the number of atoms per GPU remains fixed, the total system size increases proportionally to the number of GPUs, demonstrating weak scaling. Conversely, when the same system is distributed over more GPUs, the time required for a single MC step decreases, enabling more steps to be simulated within the same wall-clock time, an indication of strong scaling. Therefore, the strong and weak scaling behaviors in lattice parallelism represent the potential to push the atomistic simulation to larger spatial and temporal scales, respectively, by simply adding more computational resources.

The temperature parallelism can be discussed in two different situations. First, note that in the canonical ensemble, temperature is fixed, and simulations at different temperatures are essentially independent. Therefore, the MC simulation is embarrassingly parallel when properties at different temperatures are needed, such as to search for an order-disorder transition temperature. On the other hand, it is well known that the convergence speed of conventional MC can be slow, particularly near second-order transition, and replica-exchange (RE, also known as parallel tempering) is a widely used technique to address this problem. In RE, multiple replicas of the same system are simulated concurrently at different temperatures. Configurations between replicas are periodically exchanged according to a Metropolis criterion that preserves detailed balance. 
The transition probability for two configurations with energies $E_i$ and $E_j$ is given by:
\begin{equation}
    P_\text{swap} = \rm{min}(1, \exp(\Delta)),
\end{equation}
where
\begin{equation}
    \Delta = \frac{1}{k_B}\left(\frac{1}{T_i} - \frac{1}{T_j} \right)(E_i - E_j),
\end{equation}
$T_i$ and $T_j$ are the temperatures for configuration $i$ and $j$. By exchanging configurations between neighboring temperatures, the replica-exchange algorithm mitigates the risk of the system becoming trapped in local energy minima. This is especially critical near second-order phase transition points \cite{PhysRevLett.58.86, PhysRevLett.86.2050}, where the divergence of the correlation length makes it challenging for simulations at a fixed temperature to escape metastable states. For large systems, an additional complication needs to be considered: the energy fluctuation decreases as the system size increases. Consequently, the energy distributions of adjacent replicas may overlap insufficiently, leading to low exchange rates. To ensure reasonable exchange rate, the temperature difference $\Delta T$ need to be adjusted in accordance with the system size $N_{\text{atom}}$ as:
\begin{equation}
    \Delta T \propto 1/\sqrt{N_{\text{atom}}},
\end{equation}
For large atomic systems, the required temperature spacing 
$\Delta T$ becomes very small (e.g., 0.1 K for 1 billion atoms), which significantly increases the number of replicas and thus the overall computational cost. Limited by computing resources, we do not attempt a full RE simulation in this work. Note that the $\rm{Fe_{29}Co_{29}Ni_{28}Al_7Ti_7}$ HEA studied in this work does not show obvious order-disorder transition in its $C_v$ curve, as illustrated in Ref.~\cite{liuNPJ2025}.

\subsection{SMC-GPU and SMC-CPU}
\begin{figure}[ht!]
  \centering
  \includegraphics[width=1. \linewidth]{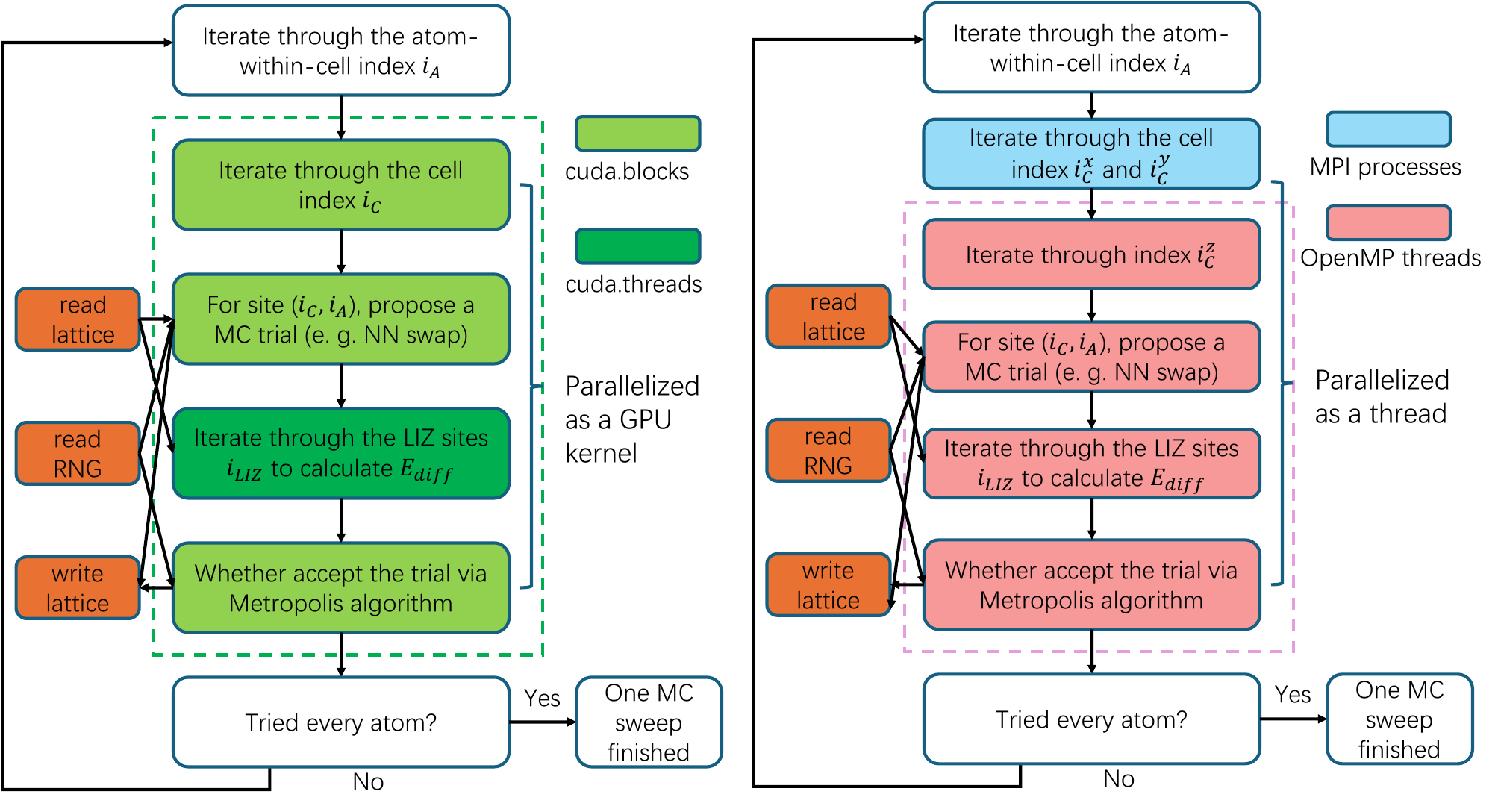}
  \caption{Illustration of the SMC-GPU and SMC-CPU algorithms. In SMC-GPU, the LIZ and LC degree of parallelism are allocated to the \textit{cuda.blocks} and \textit{cuda.threads}. In SMC-CPU, part of the LC degree of parallelism (x and y directions) are parallelized via MPI processes, while the other LC degree of parallelism, as well as the LIZ ones, are parallelized via OpenMP threads.} \label{fig:GPU+CPU}
\end{figure}

As shown in Fig.~\ref{fig:parallelism}, the SMC-X framework contains a hierarchy of algorithmic parallelisms. To efficiently utilize these opportunities, the various algorithmic parallelism need to be carefully mapped into the hardware parallelism in modern high-performance computers, in which the main computing capability typically coming from CPUs or GPUs. Considering the differences in CPU and GPU, such as hardware architecture and programming model, we designed different variants of SMC-X: SMC-GPU and SMC-CPU, as illustrated in Fig.~\ref{fig:GPU+CPU}. 

On CPUs, fine-grained tasks such as ML model evaluations and local energy updates are accelerated by vector and matrix units, while medium-grained tasks such as link-cell and lattice decomposition are distributed over CPU cores and intranode processors using SIMD, OpenMP, and MPI programming. Coarse-grained tasks, such as temperature parallelism via replica exchange, are mapped across computing nodes. On GPUs, an analogous hierarchy is employed: CUDA and tensor cores handle fine-grained ML operations, streaming processors execute local interaction and link-cell computations, and intranode GPUs coordinate lattice decomposition. Finally, replica exchange across temperatures is distributed over GPU computing nodes via CUDA/MPI. This hierarchical mapping strategy ensures that algorithmic parallelism is effectively matched to the granularity of available hardware parallelism, thereby maximizing scalability and efficiency of SMC-X simulations.

\label{Method:model}

\section{Results and discussion}
In this section, we present the simulation results for the $\rm{Fe_{29}Co_{29}Ni_{28}Al_7Ti_7}$ HEA. Two models, EPI and qSRO, are used. The EPI model\cite{LIU2021110135, ZHANG2020108247} contains only pairwise interactions, which is computationally highly efficient since there is no need to explicitly evaluate the local energy changes within the LIZ. On the other hand, the qSRO model \cite{liuNPJ2025} represents a general nonlinear short-range model, where the local energies of all the lattice sites within LIZ are calculated to determine the energy change due to a MC move. For SMC-GPU, the computational platform is a 2-node GPU cluster. Each computational node contains 8 80-GB NVIDIA H800 GPUs interconnected via 400 GB/s NVLink, and the 2 nodes are connected with a 200 GB/s InfiniBand network. For SMC-CPU, the platform is a node with 2 Kungpeng ARM CPUs. Each CPU contains multiple NUMA domains, with dozens of physical cores in each domain. 

The DFT data-set is calculated with the LSMS method \cite{PhysRevLett.75.2867}. LSMS is an all-electron electronic structure calculation method in which the computational cost scales linearly with respect to the number of atoms. For FeCoNiAlTi, we use a 100 atom supercell with a lattice constant of 6.72 Bohr (0.356 nm). We employ a spin-polarized scheme to account for the magnetic interactions in the system. The angular momentum cutoff $l_{max}$ for the electron wavefunctions is chosen as 3, and the LIZ is chosen as 86 atoms. We used PBE as the exchange-correlational functional. To enhance the representativeness of the DFT data, the total dataset is made up of 120 randomly generated configurations and 100 configurations from Monte Carlo simulation, as proposed in~\cite{LIU2021110135}. Therefore, the total number of configurations is 220, with each comprising 100 atoms. 
\subsection{Performance}

\begin{figure}[ht!]
  \centering
  \begin{subfigure}{0.48\textwidth}
    \centering
    \includegraphics[width=\linewidth]{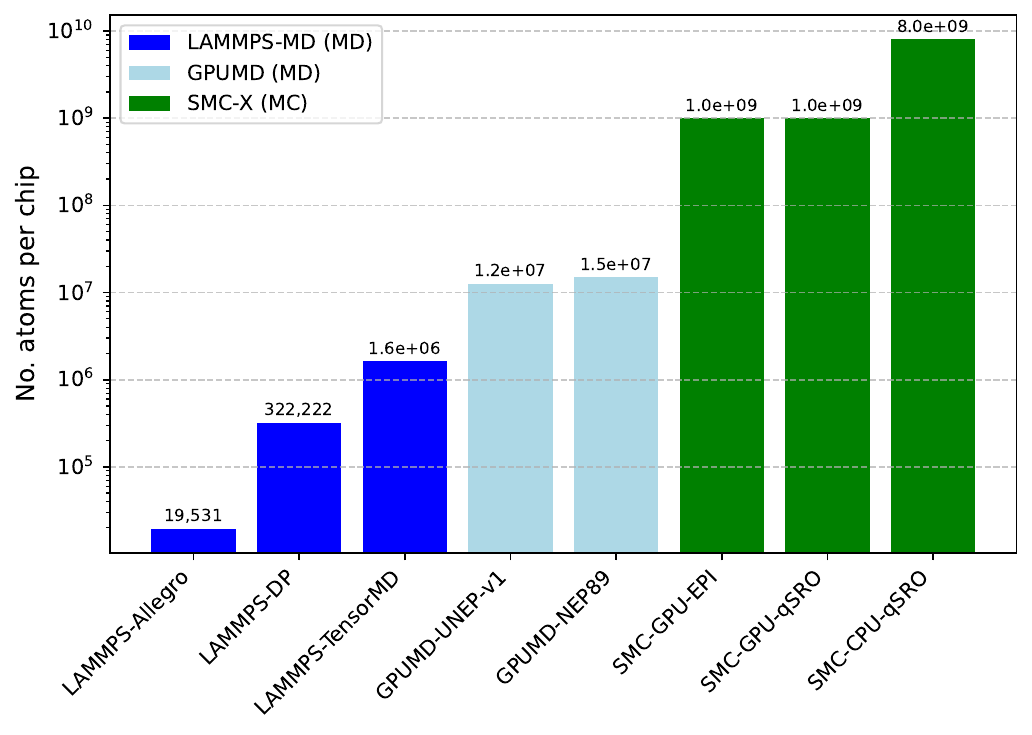}
    \caption{\add{Atoms per chip comparison}}
    \label{fig:atoms_per_chip}
  \end{subfigure}
  \hfill
  \begin{subfigure}{0.48\textwidth}
    \centering
    \includegraphics[width=\linewidth]{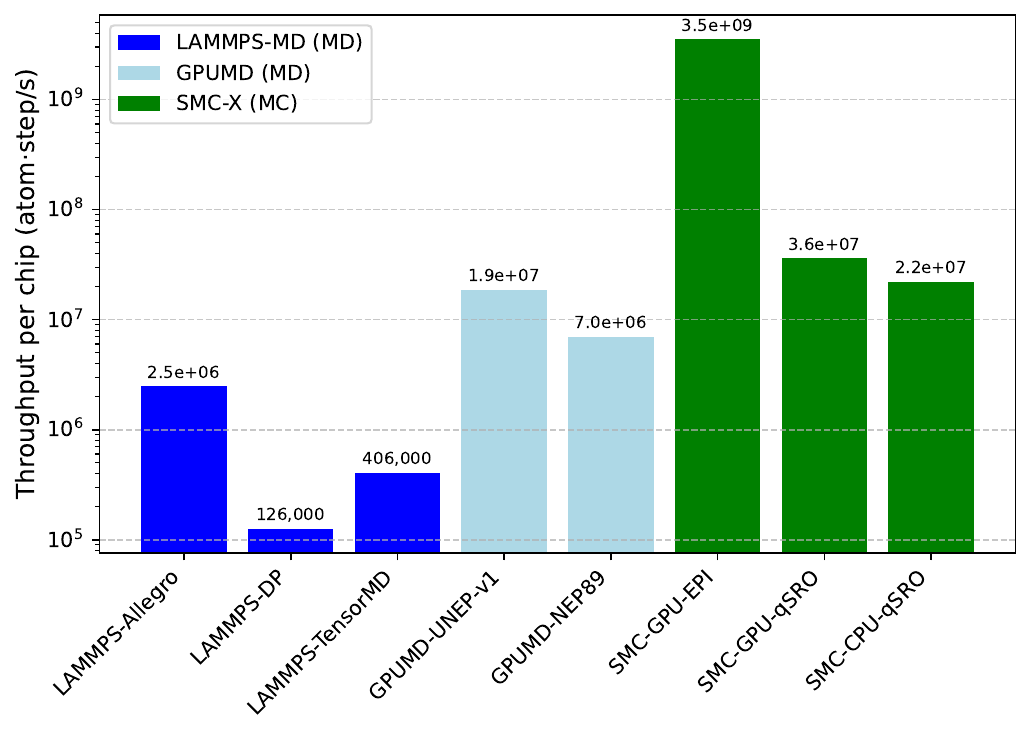}
    \caption{\add{Throughput per chip comparison}}
    \label{fig:tp_per_chip}
  \end{subfigure}
  
  \vspace{0.5cm} 
  
  \begin{subfigure}{0.7\textwidth}
    \centering
    \includegraphics[width=\linewidth]{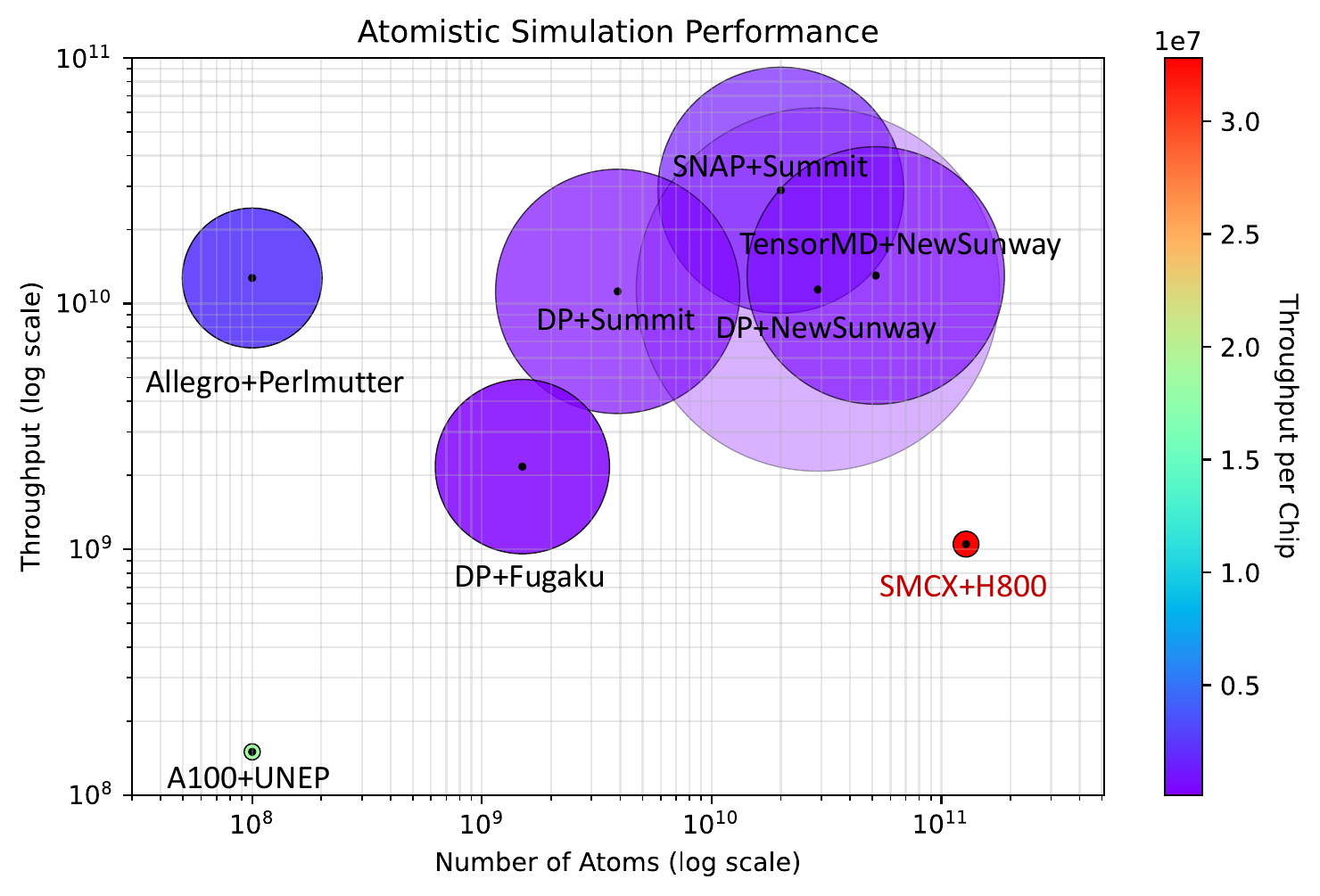}
    \caption{\add{Overall performance metrics}}
    \label{fig:jctc_performance}
  \end{subfigure}
  
  \caption{\add{Comparison of different ML accelerated atomistic simulation methods: 
           (a) shows the number of atoms that can be simulated per chip, 
           (b) displays the computational throughput achievable per chip, and 
           (c) comparison of the performance of our results with previous SOTA in terms of total $N_{atom}$ ($x$-axis),  total throughput ($y$-axis), throughput per chip (color), and the number of chips (circles with radius proportional to $N_{Chip}^{1/3}$). See original data at Tab.~\ref{tab:Performance}.}}
  \label{fig:Metrics}
\end{figure}

\begin{table}[h!]
  \caption{Comparison of the performance of atomistic simulations achieving DFT-level accuracy (10 meV/atom). Throughput (TP) is quantified as $N_\text{atom}\times N_\text{step}/T$, where $T$ is the simulation time in seconds, with $\text{atom}\cdot \text{step}/\text{s}$ as the unit. DP stands for DeepPotential.
  }
  \label{tab:Performance}
\begin{tabular}{c c c c c c c c}
\toprule
\textbf{Simu.} & \textbf{Potential} & \textbf{Sys.} & \textbf{Chips} & \textbf{Machine} & \textbf{Atoms} & \textbf{TP} & \textbf{TP/Chip} \\
\midrule
LAMMPS \cite{10.1145/3581784.3627041}& Allegro & HIV & 5,120 A100& Perlmutter & 100\;M & $1.27 \times 10^{10}$ & ${2.47 \times 10^{6}}$ \\
LAMMPS \cite{10880101}& DP & H\textsubscript{2}O & 90,000 CPU& Sunway & 29\;B & ${1.14 \times 10^{10}}$ & $1.26 \times 10^{5}$ \\
\add{LAMMPS \cite{10.1145/3503221.3508425} } & DP & H\textsubscript{2}O &  27,360 V100 & Summit & 3.9B & $1.12 \times 10^{10}$ & $4.11 \times 10^{5}$ \\
\add{LAMMPS \cite{10.1145/3503221.3508425}}  & DP & H\textsubscript{2}O & 9,936 CPU & Fugaku & 1.5B & $2.17 \times 10^{9}$ & $2.18 \times 10^{5}$ \\
\add{LAMMPS \cite{10.1145/3458817.3487400} } & SNAP & C & 27,900 V100 & Summit & 20B & $ \mathbf{2.89 \times 10^{10}}$ & $1.08 \times 10^{6}$ \\
LAMMPS \cite{10.1145/3710848.3710882}& TensorMD & W & 32,000 CPU & Sunway & \textbf{51.8\;B} &  ${{1.30 \times 10^{10}}}$ & $4.06 \times 10^{5}$ \\ 
GPUMD \cite{MLPAlloys2024}& UNEP-v1 & Cu & 8 A100  & 1 node & 100\;M & ${1.50 \times 10^{8}}$ & $1.86 \times 10^{7}$ \\
GPUMD \cite{liang2025nep89universalneuroevolutionpotential}& NEP89 & HEA & 1 H800  & 1 node & 15\;M & ${7.0 \times 10^{6}}$ & $7.0 \times 10^{6}$ \\\hline
SMC-X& EPI & HEA & 16 H800 & 2-node & {16\;B} & $\mathbf{5.63 \times 10^{10}}$ & $\mathbf{3.52 \times 10^{9}}$ \\
SMC-X  & qSRO & HEA & 16 H800 & 2-node  & {16\;B} & $5.79 \times 10^{8}$ & ${\mathbf{3.62 \times 10^{7}}}$ \\
\add{SMC-X} & \add{qSRO} & \add{HEA} & \add{32 H800} & \add{4-node} & \add{\textbf{128\;B}} & \add{$1.17 \times 10^{9}$} & \add{$3.29 \times 10^{7}$} \\
SMC-X  & qSRO & HEA & 2 CPU & 1 node & 16\;B & $4.44\times 10^{7}$ & ${{2.22 \times 10^{7}}}$ \\
\bottomrule
\end{tabular} 
\end{table} 

Performance is measured with three metrics: number of atoms, which measure spatial scale; simulation throughput, which measures time to solution; and scaling efficiency, which measures the scaling potential with increasing computational resources. As shown in \add{ Fig.~\ref{fig:tp_per_chip}}, the EPI model exhibits exceptional computational efficiency on a single GPU, achieving a throughput of $3.5\times10^9$ $\text{atom}\cdot\text{step}/\text{s}$—two orders of magnitude faster than the qSRO\_GPU result of $3.7\times10^7$ $\text{atom}\cdot\text{step}/\text{s}$. In comparison, the single-NUMA performance of qSRO\_CPU reaches $3.5\times10^6$ $\text{atom}\cdot\text{step}/\text{s}$, or $2.5\times10^7$ $\text{atom}\cdot\text{step}/\text{s}$ per CPU. The throughput of SMC-X not only far exceeds that of traditional Monte Carlo methods, but also outperforms state-of-the-art molecular dynamics implementations, as shown in Fig.~\ref{fig:Metrics} and summarized in Tab.~\ref{tab:Performance}. \add{It is noteworthy that SMC-X, utilizing merely 32 H800 GPUs, can simulate systems of 128 billion atoms, which is a scale that surpasses all other ML-accelerated MC or MD simulations, including those performed on some of the world's most powerful supercomputers, as shown in Fig.~\ref{fig:jctc_performance}.}
The advantage is even more pronounced in terms of the number of atoms supported on a single chip: the SMC-GPU implementation accommodates up to \add{4 billion atoms on a single GPU by running 4 MPI ranks simultaneously}, whereas the best reported MD simulation to date reaches only 15 million atoms per GPU, as shown in the row of NEP89 in Tab.~\ref{tab:Performance}. As mentioned, these performance gains stem from key implementation optimizations, including on-the-fly computation and tight coupling between atomistic simulation and machine learning models, which collectively minimize data movement and maximize computational efficiency.

\begin{figure}[ht!]
  \centering
  \includegraphics[width=0.7\linewidth]{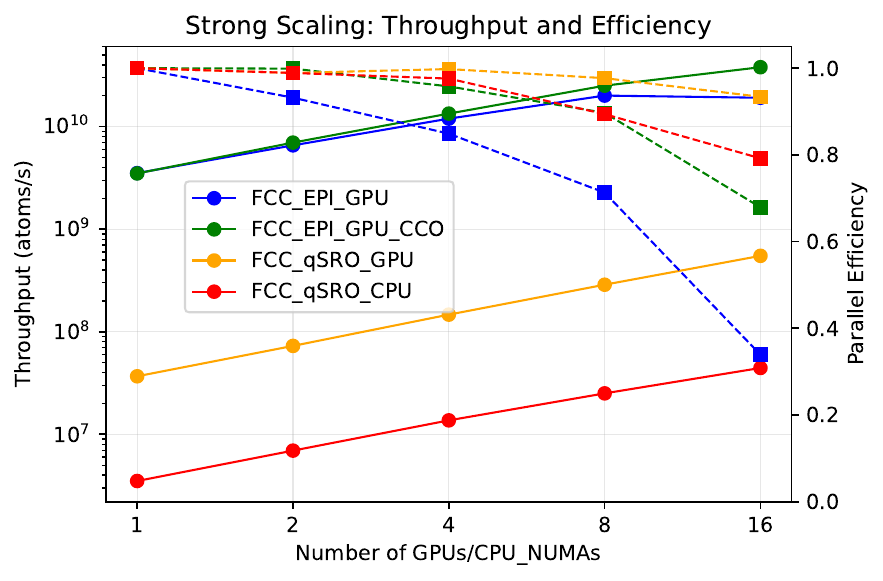}
  \includegraphics[width=0.7\linewidth]{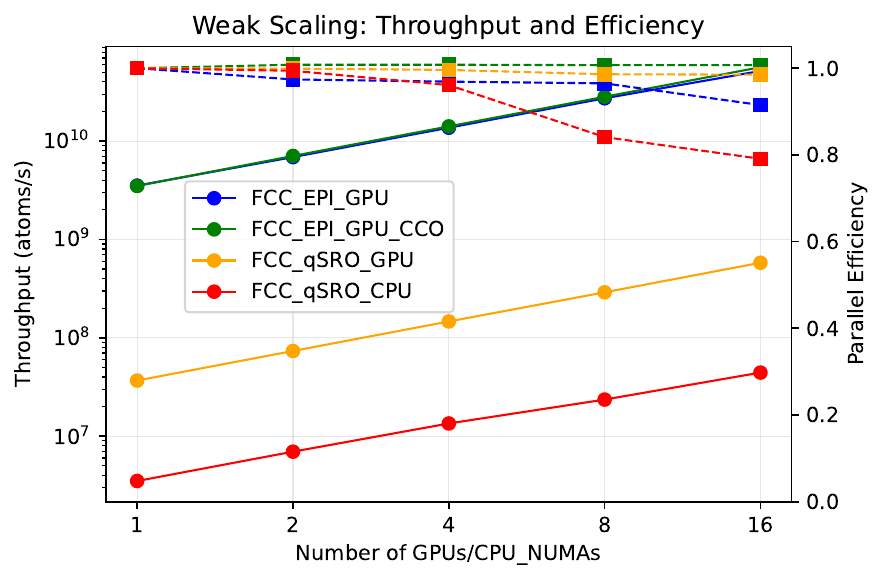}
  \caption{The strong and weak scaling of SMC-GPU and SMC-CPU with EPI and qSRO model. CCO represents computation-communication-overlap.} \label{fig:Scaling}
\end{figure}
The strong and weak scaling efficiencies of various methods are shown in Fig.~\ref{fig:Scaling}. We also implemented a computing-communication overlap for EPI\_GPU, as efforts to enhance the scaling efficiency for communication heavy cases. The results of all the above cases are shown in Fig.~\ref{fig:Scaling}.
First, we find that the qSRO model consistently demonstrates superior scaling efficiency due to its higher computation-to-communication ratio. For example, the strong scaling efficiency of qSRO\_GPU reaches 0.934 on 16 GPUs, compared to only 0.339 for EPI\_GPU. Second, the strong scaling of EPI\_GPU can be substantially improved from 0.339 to 0.679 on 16 GPUs by applying computation–communication overlap (CCO). Notably, on 8 GPUs, CCO enhances efficiency from 0.713 to 0.897. These results highlight that CCO is highly effective in improving scaling behavior, particularly when combined with high-speed interconnects (e.g., Nvidia NVLINK within 8 GPUs) and large system sizes.
Third, all methods exhibit excellent weak scaling, with 16-GPU efficiencies of 0.915, 1.007, 0.985, and 0.791 for EPI\_GPU, EPI\_GPU\_CCO, qSRO\_GPU, and qSRO\_CPU, respectively. This demonstrates that SMC-X can be readily extended to larger spatial scales. Finally, although both strong and weak scaling efficiencies for qSRO\_CPU are generally good, they are consistently lower than their GPU counterparts, indicating that cross-CPU NUMA communication is more costly than GPU interconnects such as NVLINK.

\add{Finally, we comment on the limitations of current SMC-X implementations and its future potential. The present implementation requires kernel-level programming to optimize the machine-learning (ML) potential, which enables peak performance but makes the integration of additional potentials labor-intensive. Consequently, we have so far implemented only a simple qSRO model as a representative of general short-range models. This naturally raises the question of whether such performance can be sustained with more complex ML interatomic potentials (MLIPs). To investigate this, we performed a roofline analysis of the SMC-GPU code using NVIDIA Nsight tools. The results show that the current implementation is bounded primarily by integer arithmetic for on-the-fly index calculations and by HBM bandwidth for data movement, whereas the floating-point units, including CUDA vector units and Tensor Cores, remain significantly underutilized due to the low theoretical arithmetic intensity of about 1. Given that modern GPUs are designed for workloads with arithmetic intensities in the hundreds or higher, it is reasonable to expect that incorporating more complex models with higher computational intensity will not compromise performance. Indeed, in a prototype implementation using a three-layer multilayer perceptron (MLP) with a hidden dimension of 256, we observe that the excellent speed reported for qSRO is retained. Looking ahead, accommodating fully pretrained MLIPs may require a modest redesign of the SMC-X code architecture, potentially introducing a small performance overhead. Nevertheless, the method would remain exceptionally fast compared with conventional Monte Carlo approaches, as the dominant speedup arises from the SMC-X algorithm itself. These enhancements will be explored in future work.}

\subsection{Evolution of chemical defects}

\begin{figure}[ht!]
  \centering
  \begin{subfigure}{0.48\textwidth}
    \centering
    \includegraphics[width=\linewidth]{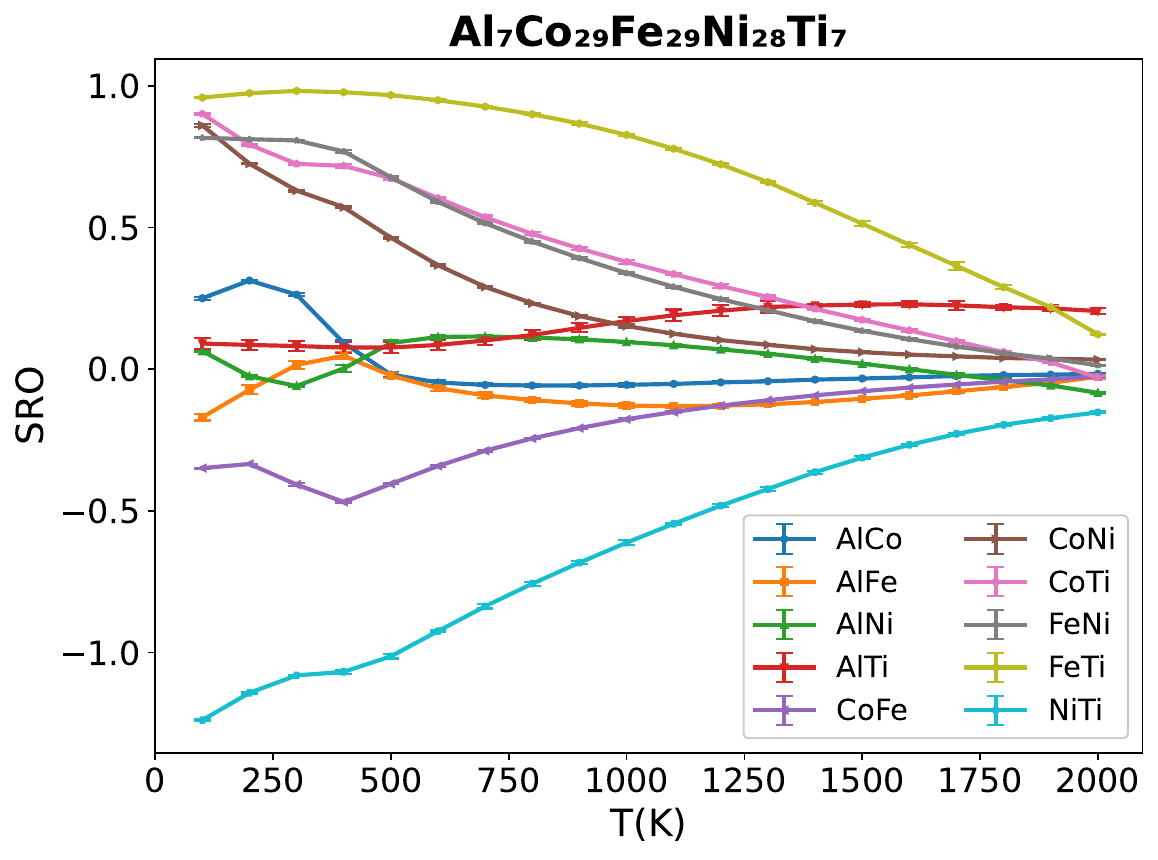}
    \caption{\add{SRO of different element pairs}}
    \label{fig:atoms_per_chip}
  \end{subfigure}
  \hfill
  \begin{subfigure}{0.48\textwidth}
    \centering
    \includegraphics[width=\linewidth]{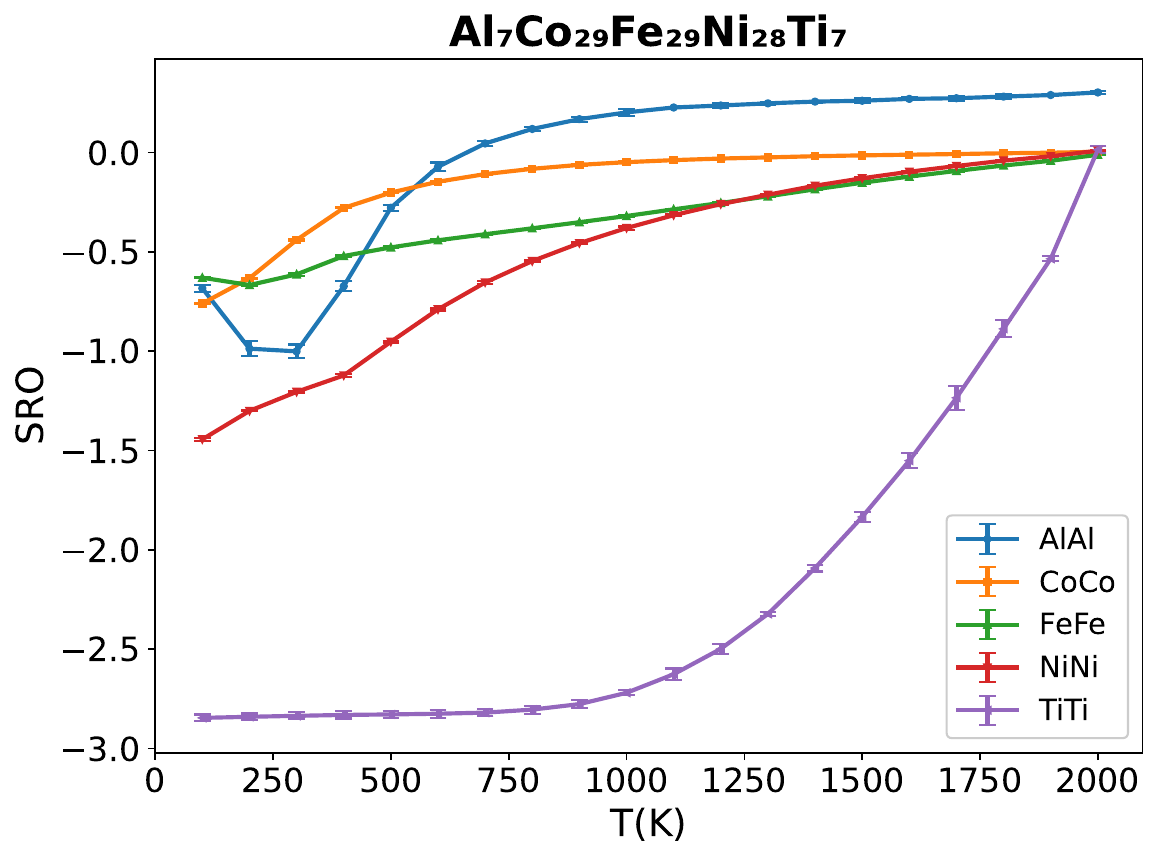}
    \caption{\add{SRO of identical element pairs}}
    \label{fig:tp_per_chip}
  \end{subfigure}

  \caption{The chemical short-range order parameters (SROs) calculated with the 1 million atom supercell, \add{estimated with 9 configuration samples. The error bars are estimated with the difference between the largest and smallest ones among the 9 SRO values.}(a) The SROs for different elements pairs; (b) The SROs for the same element pairs.}
  \label{fig:SRO}
\end{figure}

The Warren--Cowley short-range order (SRO) parameter \cite{PhysRev.77.669, LIU2021110135}, denoted $\alpha_m^{pp'}$, is widely used to quantify local chemical ordering. It is defined as
\begin{equation}
\alpha_m^{pp'} = 1 - \frac{P^{p|p'}_m}{c_p}, 
\label{SRO}
\end{equation}
where $c_p$ is the concentration of species $p$, and $P^{p|p'}_m$ is the conditional probability of finding an atom of type $p$ in the $m$-th neighbor shell around a central atom of type $p'$. The SRO parameters are obtained by averaging over the Monte Carlo configurations. By definition, $\alpha=0$ corresponds to perfect randomness, $\alpha<0$ indicates a tendency toward ordering between $p$ and $p'$, and $\alpha>0$ reflects a tendency toward segregation.
Figure~\ref{fig:SRO} shows the nearest-neighbor (NN) SRO parameters obtained from Monte Carlo simulations of a one-million-atom FeCoNiAlTi system. At high temperature, the alloy remains nearly random, whereas at lower temperatures distinct SRO develops. In particular, Ti and Ni atoms display a strong tendency to segregate from Fe and Co, with the Ti--Fe pair showing the strongest separation. This behavior is consistent with previous experimental \cite{Yang933} and theoretical \cite{liuNPJ2025} studies, which reported the formation of $L1_2$-type nanoprecipitates enriched in Ni and Ti within a Fe- and Co-rich matrix.

\begin{figure} [ht!]
    \centering
    \includegraphics[width=1. \linewidth]{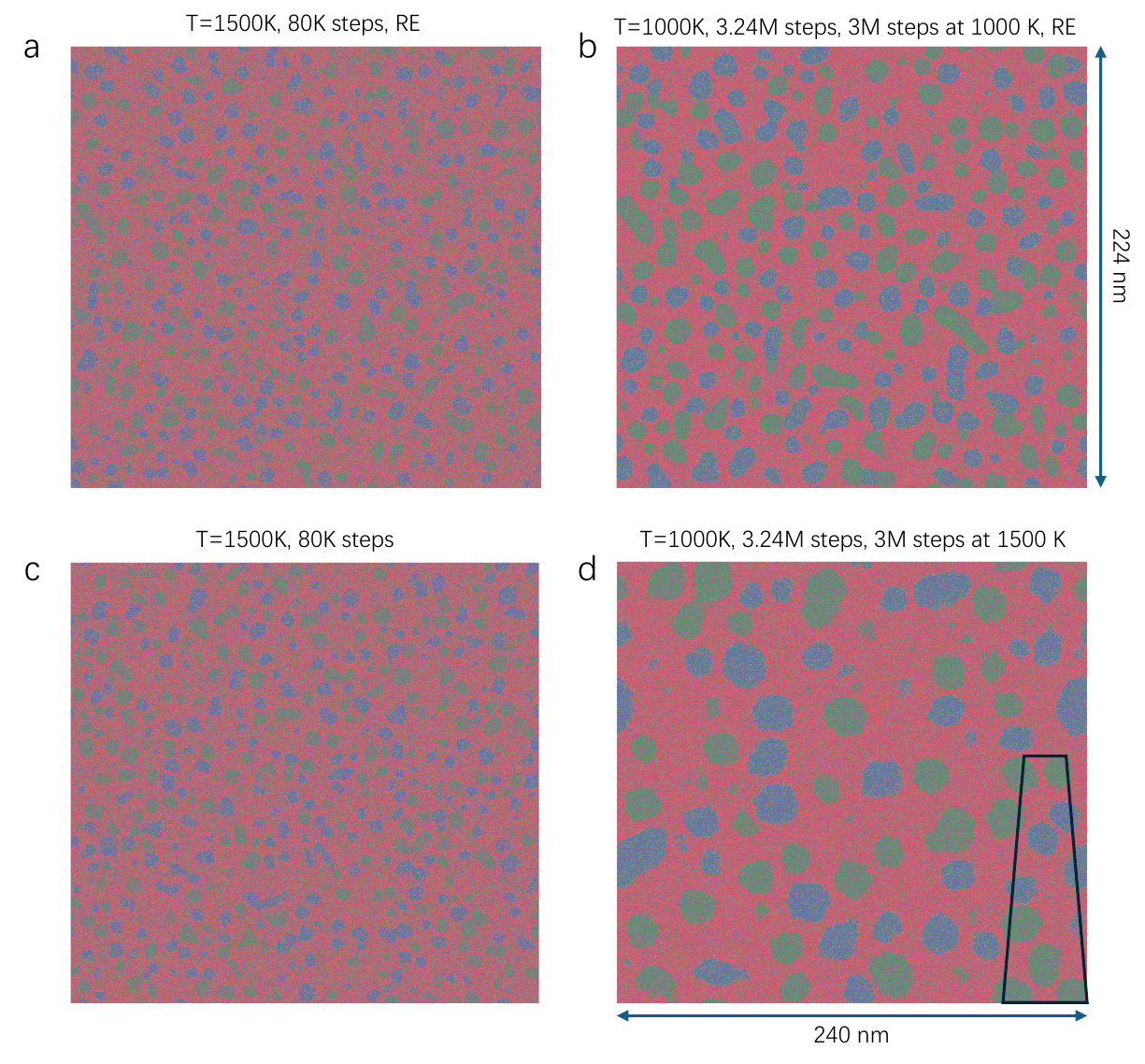}
    \caption{Simulation results at different temperature schedule for 1-billion-atom $\rm{Fe_{29}Co_{29}Ni_{28}Al_7Ti_7}$. Note that two FCC layers are shown in a checkerboard grid for clarity of the image. (a-d) The X-Y faces at different stages of the MC simulation, as denoted in the subtitles. Atoms are colored as Fe: red; Co: purple; Ni: green; Al: grey; Ti: blue.}
    \label{fig:1B_Box}
\end{figure}

To investigate microstructure evolution,
we performed large-scale Monte Carlo simulations using the EPI model, with a supercell comprising 1,066,867,200 atoms (dimensions: $672 \times 630 \times 630 \times 4$). 
The computing platform is the two-node, 16-H800 GPU cluster, and the ML model is the EPI model to maximize the simulation speed. 
We decrease the temperature from 2000 K to 1000 K, with tens of thousands of MC steps at each temperature, except for a temperature we refer to as the ``aging" temperature $T_{aging}$, at which the simulation runs for 3 million MC steps . The X-Y face of the system for $T_{aging}=1000$ K experiment is shown in Fig.~\ref{fig:1B_Box} (a, b), where we used 8 GPUs for lattice parallelism and 2 temperature degrees of parallelism, with $\Delta T=0.06$ K. As mentioned in the Methods section, such a small $\Delta T$ is for a reasonable acceptance ratio of RE in large supercells. The results for $T_{aging}=1500$ K is shown in Fig.~\ref{fig:1B_Box} (c, d), where no RE is used, and the lattice is parallelized over the total 16 GPUs. From these figures, it can be seen that the NPs rich in Ni and Ti already form at 1500 K. Upon further cooling, nanoprecipitates gradually grow in size. Moreover, the comparison between Fig.~\ref{fig:1B_Box} (a) and (c) suggests that the two temperature replicas demonstrate very little effect, which is consistent with our expectation, due to the tiny value of $\Delta T$. Finally, it can be seen that the size of NPs increases as $T_{aging}$ switched from 1000 K to 1500 K, indicating the high rate of NP formation at elevated temperature.

\subsection{APT sample}
\begin{figure} [htbp]
    \centering
    \includegraphics[width=1. \linewidth]{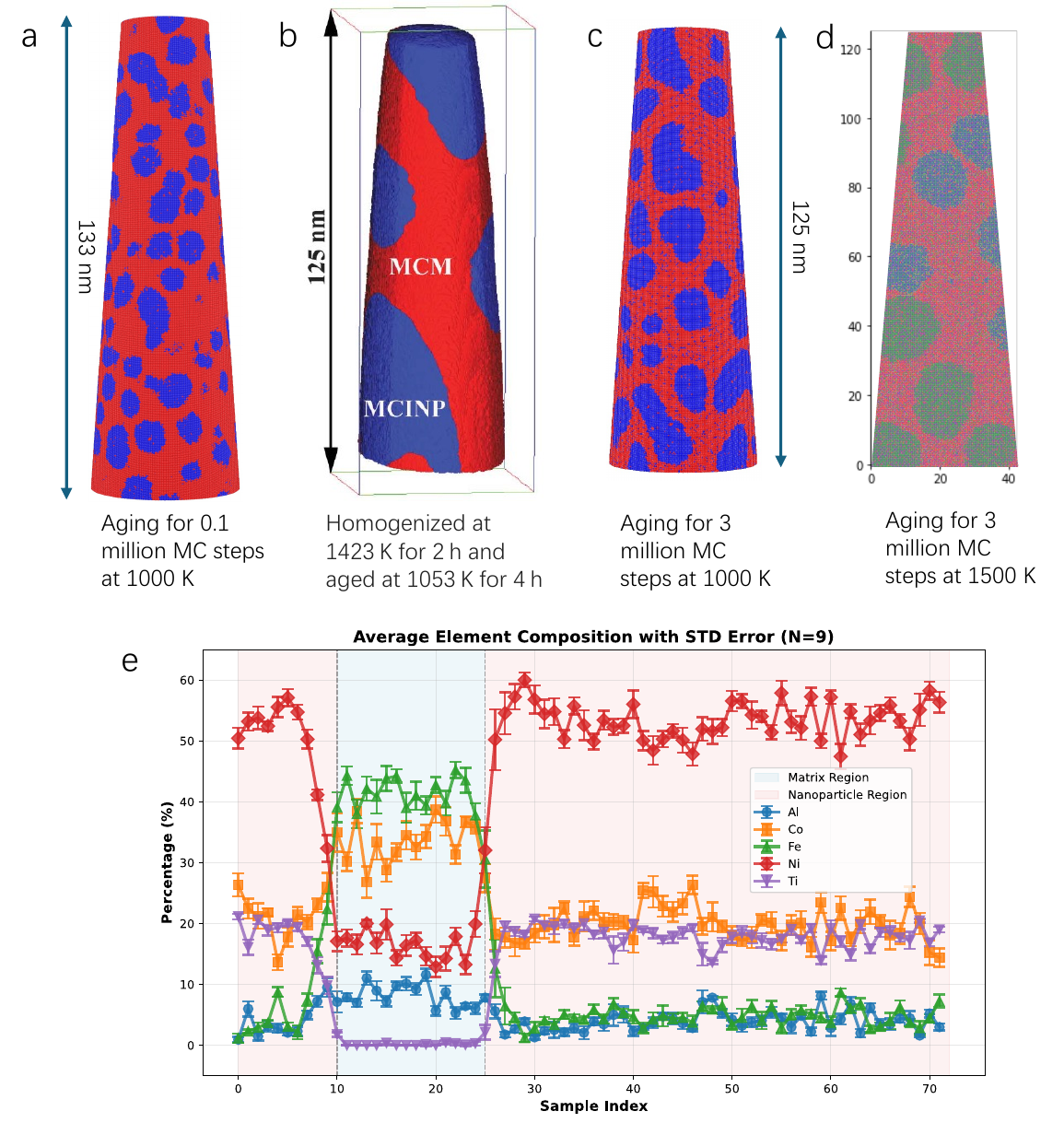}
    \caption{Comparison of APT sample from the simulation with previous experimental and theoretic studies. (a) The APT sample extracted from SMC-X simulation in Ref.~\cite{liuNPJ2025}. \add{Reproduced from Ref.~\cite{liuNPJ2025}. Available under a CC-BY NC-ND 4.0 license. Copyright Xianglin Liu et al. }(b) The APT reconstructed sample image of the same materials from experiment \cite{Yang933}, where MCM represents multicomponent matrix, and MCINP represents multicomponent intermetallic nanoparticle. \add{Reproduced with permission from Ref.~\cite{Yang933}. Copyright 2018 AAAS.} (c) The APT sample from this work, using 3 million MC steps at 1000 K. (d) The APT sample extracted from Fig.~\ref{fig:1B_Box}\ d, using 3 million MC steps at 1500 K. (a-c) blue represents MCINP, and red represents MCM. (e) Averaged compositions obtained from 9 nearby cylinderical specimen of $r=12\;a$ and $l=72\;a$, where $a$ is the lattice constant.}
    \label{fig:APT}
\end{figure}

To reveal further details of the NPs, we extract an atom-probe-tomography (APT) sample from the simulation results and compare them with previous experimental and theoretical studies, as shown in Fig.~\ref{fig:APT}. First, we see that the previous study shown in Fig.~\ref{fig:APT}(a) clearly underestimates the sizes of the NPs, which is about 7 nm in diameter, compared to the experimental result in Fig.~\ref{fig:APT} (b), which is about 30 nm. Second, we see that this gap between experiment and theory can be reduced by increasing the MC steps from 0.1 million to 3 million, which increases the size of the NP to about 14 nm. 
Finally, note that the experimental sample is synthesized by homogenizing at 1423 K for 2 h and aged at 1053 K for 4 h. The 3 million steps at 1500 K in Fig.~\ref{fig:APT}(d) can be seen as an effort to mimic the homogenization effect. It can be seen that the size of the NP can be further increased to about 20 nm.
These observations reveal that a larger number of steps are needed for the system to reach equilibrium, and the discrepancy in the size of the nanoparticles in Ref.~\cite{Yang933}
and Ref.~\cite{liuNPJ2025} can be narrowed by increasing the simulation steps to mimic the whole heat-treatment process. We emphasize that the distributed implementation presented in this is the reason that we can run such a large-number of MC steps in a time budget of a few days. \add{The chemical composition at 1000 K is analyzed, as shown in Fig~\ref{fig:APT} (e). The composition profile is extracted from 9 nearby cylindrical specimen of radius $r=12\;a$ and length of $l= 72 \;a$, where $a$ is the lattice constant. It can be seen that Fe and Co are favored in the disordered matrix phase, while Ni and Ti and strongly favored in the ordered (Ni,Co)3-(Ti,Al) $L1_2$ phase. These results agrees well with both experiments  \cite{Yang933} and previous theoretical studies \cite{liuNPJ2025, WANG2025120635}. The error bar is the standard deviation of the compositions.}

Although there is no time dependence in the MC method, the time scale of the simulation can be estimated using a simple diffusion model, similar to kinetic MC. A well-known model to describe the diffusion-controlled coarsening is Lifshitz–Slyozov–Wagner (LSW) theory, which gives 
\begin{align}
    r(t)^3-r(0)^3=Kt
\end{align}
where $r(t)$ is the average precipitate radius at time $t$, and $K$ is the diffusion speed, which depends exponentially on the energy barrier for swapping the atoms. Note that the experimental diameter of the NP is approximately 30 nm, and it takes about 2.9 M MC steps to grow from 7 nm to 14 nm at T=1000 K, therefore, the number of MC steps needed to reach the experimental size is
\begin{align}
    n_{\text{exp}} = \frac{30^3}{(14^3-7^3)}\times\left(3\times 10^6-10^5\right)= 3.26\times10^7
\end{align}
Using the speed of $5\times10^{10}$ atom$\cdot$step/s for a 16 GPU system, then it would have taken approximately 75 days. Limited by computing resource, we did not run this full simulation, but we argue that this is an achievable goal, particularly with speedup from further optimization on larger computing platform. On the other hand, the aging process in alloys typically takes hours. Assuming the NP in experiment is formed in 1 hour, then the physical time corresponds to in our 3M step simulation is 
\begin{align}
    t_{\text{sim}}=\frac{3600\times 3\times10^6}{3.26\times10^7} s =331\;s 
\end{align} 
Of course, this is a very crude estimation, but it indicates that our MC simulation indeed reaches the time regime of a real physical process of heat-treatment.

\section{Conclusions}
In this work, we have presented a distributed framework for the SMC-X method, demonstrating its versatility with implementations on both CPU and GPU architectures. Using hierarchical parallelism, data-locality optimization, and computation–communication overlap, the distributed SMC-X achieves exceptional performance in simulation throughput and scaling efficiency. \add{In particular, it achieves the largest system size for ML-accelerated atomistic simulation, reaching 128 billion atoms using only 32 H800 GPUs, a much smaller hardware budget compared to previous results obtained with the largest supercomputers.} When combined with DFT-trained machine learning models, SMC-X attains first-principles accuracy while extending both the spatial and temporal scales of atomistic simulations to previously inaccessible regimes. Such capability is essential for predicting large-scale chemical defects in chemically complex alloys and, in turn, for elucidating the origin of their exceptional properties. This claim is exemplified by our billion-atom simulations of FeCoNiAlTi, where extended spatial and temporal scales are critical for bridging the gap between experimental observations and theoretical predictions. Together, our results demonstrate that the SMC-X + ML framework sheds light on the challenging but critical problem of simulation-driven design of chemically complex materials.

\section*{Data Availability}  
The data and analysis code supporting the findings of this study are available at: \url{https://github.com/xianglil/JCTC-SMCX-Supplementary}

\section*{Code Availability}  
A binary version of the code is available upon reasonable request. 

\section*{Author contributions}
X.L. conceived and supervised the project, developed the code prototype, fitted the ML model, and wrote the original manuscript. K.Y. implemented the production code and collected the original data. F.Z. performed data analysis and visualization. P.X. acquired and managed the computing resources. All authors discussed the results and contributed to the final version of the manuscript.

\begin{acknowledgement}
The work of X. Liu and F. Zhou was supported by the National Natural Science Foundation of China under Grant 12404283. \add{The work of F. Zhou was also supported by the High-Level Talent Start-up Fund provided by Xiangnan University. }X. Liu acknowledges helpful discussion with Zongrui Pei on precipitate sizes in alloy.
\end{acknowledgement}

\section*{Competing Interests}
The authors declare no conflict of interest.

\bibliography{achemso-demo}

\end{document}